\documentstyle[aasms4]{article}
\begin{document}

\title{Solar System Tests of Higher-Dimensional Gravity}

\author{Hongya Liu}
\affil{Department of Physics, Dalian University of Technology,
       Dalian 116024, P. R. China}
\and
\author{J. M. Overduin}
\affil{Department of Physics, University of Waterloo,
       Waterloo, ON, Canada N2L 3G1}
\affil{and}
\affil{Gravity Probe~B, Hansen Experimental Physics Laboratory,
       Stanford University}

\begin{abstract}
The classical tests of general relativity --- light deflection, time delay
and perihelion shift --- are applied, along with the geodetic precession test,
to the five-dimensional extension of the theory known as Kaluza-Klein gravity,
using an analogue of the four-dimensional Schwarzschild metric.
The perihelion advance and geodetic precession calculations are generalized
for the first time to situations in which the components of momentum and 
spin along the extra coordinate do not vanish.
Existing data on light-bending around the Sun using long-baseline
radio interferometry, ranging to Mars using the Viking lander, and the
perihelion precession of Mercury all constrain a small parameter $b$
associated with the extra part of the metric to be less than $|b| < 0.07$ 
in the solar system.  An order-of-magnitude increase in sensitivity
is possible from perihelion precession, if better limits on solar oblateness
become available.  Measurement of geodetic precession by the Gravity 
Probe~B satellite will improve this significantly, probing values of $b$
with an accuracy of one part in $10^4$ or more.
\end{abstract}

% \pacs{04.50, 04.80}

\section{Introduction}

There is now a substantial literature on the higher-dimensional extension
of Einstein's general theory of relativity known as Kaluza-Klein gravity
(\cite{Ove97,Wes99}).  There are several ways to test the theory,
with perhaps the most straightforward involving the motion of test particles
in the field of a static, spherically-symmetric mass like the Sun or the 
Earth.  Birkhoff's theorem in the usual sense does not hold in higher 
dimensions (\cite{Bro95,Sch97}), so some question arises in identifying the 
appropriate metric to use for this problem.  In the five-dimensional (5D) 
case (with one extra coordinate $y\equiv x^4$), most attention has
focused on the {\em soliton\/} metric (\cite{Gro83,Sor83,Dav85}),
which satisfies the 5D vacuum field equations, reduces to the standard 
four-dimensional (4D) Schwarzschild solution on hypersurfaces $y=$~const,
and contains no explicit $y$-dependence.  The assumption of a vacuum in 5D
is consistent with the spirit of Kaluza's idea, that 4D matter and gauge 
fields appear as a manifestation of pure geometry in the higher-dimensional
world.  The soliton metric has been generalized in various ways to 
incorporate time-dependence (\cite{Liu93}), $y$-dependence (\cite{Bil96}) 
and electric charge (\cite{Liu97}), among other things (eg, \cite{Wes98});
see for review Overduin and Wesson (1997).  We confine ourselves here 
to the original (two-parameter) soliton metric.

The motion of test bodies in the gravitational field of the soliton
can be studied using the familiar classical tests of general relativity
(gravitational redshift, light deflection, perihelion advance and time delay),
along with the geodetic precession test.  Work done so far along these
lines (Lim, Overduin and Wesson 1995; Kalligas, Wesson and Everitt 1995,
hereafter ``KWE'') has demonstrated the existence of small but potentially 
measurable departures from the standard 4D Einstein predictions.
In the present paper, we extend these earlier calculations in several ways,
clarifying the physical meaning of the light deflection and time delay
results for massless test particles and presenting new generalizations
of the perihelion shift and geodetic precession formulas for massive ones.
We take special care to compare our results to the latest experimental
data in each case, obtaining new numerical constraints on the 
small parameter $b$ associated with the extra part of the soliton metric.

\section{The Soliton Metric} \label{sec:SM}

In what follows,
lowercase Greek indices $\mu,\nu \ldots$ will be taken to run over $0,1,2,3$
as usual, while capital Latin indices $A,B,C \ldots$ run over all five 
coordinates ($0,1,2,3,4$).  Units are such that $G=c=1$ except where
stated otherwise.  It is important to distinguish between the 4D line 
element ($ds$) and its 5D counterpart ($dS$), the two being related by
\begin{equation}
dS^{\, 2} = ds^2 + g_{44} \, dy^2 \; .
\label{dS_ds}
\end{equation}
To interpret an expression containing $d/dS$ physically, one
can always make the conversion
\begin{equation}
\frac{d}{dS}=\frac{ds}{dS}\frac{d}{ds}=\sqrt{1-g_{44}\left(
   \frac{dy}{dS}\right)^2} \, \frac{d}{ds} \; .
\label{ddS_dds}
\end{equation}
We emphasize in particular that $d/dS\neq d/ds$ if $dy/dS \neq 0$.

The soliton metric may be written (following \cite{Gro83}, but switching
to nonisotropic form, and defining $a\equiv 1/\alpha$, $b\equiv \beta/\alpha$
and $M\equiv 2m$)
\begin{eqnarray}
dS^{\,2} \! \! \! \! & = & \! \! \! \! A^a dt^2 - A^{-a-b} dr^2 - A^{1-a-b} \, 
   \times \nonumber \\
& & \! \! \! \! \! r^2 \left( d\theta ^2 + \sin ^2\theta 
   d\phi ^2 \right) - A^b dy^2 \; , 
\label{Sol_Metric}
\end{eqnarray}
where $A(r) \equiv 1-2M/r$, $M$ is a parameter related to the mass of the 
object at the center of the geometry, and the constants $a,b$ satisfy a 
consistency relation
\begin{equation}
a^2+ab+b^2 = 1 \; ,
\label{Consist}
\end{equation}
so that any two of $M,a,b$ may be taken as independent metric
parameters.  We will treat $b$ as the primary free parameter of the 
theory in what follows, noting that the 4D Schwarzschild metric
is recovered (on hypersurfaces $y=$~const) in the limit $b \rightarrow 0$
(and $a \rightarrow +1$).  In general, larger values of $|b|$ will
give rise to increasing departures from Einstein's theory, 
subject to the upper bound $|b| \le 2/\sqrt{3} \approx 1.15$ 
imposed by equation~(\ref{Consist}).  Possible theoretical
expectations for this parameter in the solar system and elsewhere
are discussed further in \S \ref{sec:Disc}.

\section{Equation of Motion} \label{sec:EM}

We proceed now with the analysis of experimental constraints.
The Lagrangian for a test particle in the field described by 
the metric~(\ref{Sol_Metric}) is
\begin{eqnarray}
{\cal L} \! \! \! \! & = & \! \! \! \! \left[ A^a \dot{t}^2 - A^{-a-b}
   \dot{r} ^2 - A^{1-a-b} \, \times \right. \nonumber \\
 & & \! \! \! \! \left. r^2 \left( \dot{\theta }^2 + \sin ^2\theta 
   \dot{\phi } ^2\right) - A^b \dot{y}^2 \right] ^{1/2} \; ,
\end{eqnarray}
where the overdot represents differentiation with respect to an
affine parameter $\lambda$ along the geodesics.

The Euler-Lagrange equations read
\begin{equation}
\frac{d}{d\lambda }\left( \frac{\partial {\cal L}}{\partial\dot{x}^C}
   \right) -\frac{\partial {\cal L}}{\partial x^C}=0 \; .
\end{equation}
We confine ourselves to orbits with $\theta =\pi /2$ and
$\dot{\theta }=0$, so that ${\cal L}$ becomes
\begin{equation}
{\cal L}=\left( A^a\dot{t}^2-A^{-a-b}\dot{r}^2-A^{1-a-b}r^2\dot{\phi }^2-
   A^b\dot{y}^2\right) ^{1/2} \; .
\end{equation}
We can identify three constants of motion
\begin{eqnarray}
\ell \! \! \! \! & \equiv & \! \! \! \! \frac{1}{{\cal L}}\, 
   A^a\dot{t} = A^a\frac{dt}{dS} \; , \nonumber \\
h \! \! \! \! & \equiv & \! \! \! \! \frac{1}{{\cal L}} A^{1-a-b}
   r^2 \dot{\phi} = A^{1-a-b} r^2 \frac{d\phi}{dS} \; , \nonumber \\
k \! \! \! \! & \equiv & \! \! \! \! \frac{1}{{\cal L}} A^b\dot{y} =
   A^b\frac{dy}{dS} \; ,
\label{ell_h_k}
\end{eqnarray}
where we have used the relation ${\cal L}=dS/d\lambda$. From these
equations we find that
\begin{eqnarray}
\left( \frac{dr}{d\phi }\right) ^2 \! \! \! \! & + & \! \! \! \! A 
   r^2 - \left( \frac{\ell ^{\, 2}}{h^2} \, A^{2-2a-b} - \frac{k^2}{h^2} \, 
   A^{2-a-2b} \right. \nonumber \\
& - & \! \! \! \! \left. \frac{1}{h^2} \, A^{2-a-b} 
   \right) r^4=0 \; .
\label{r-DE}
\end{eqnarray}
The derivation here differs slightly from that of KWE,
where ${\cal L}\equiv (dS/d\lambda )^2$.  Although the two approaches
are physically equivalent, we have found that results are obtained more
simply if the three constants of motion $\ell,h,k$ are defined in terms
of $d/dS$ rather than $d/d\lambda$ (or $d/ds$).

\section{Light deflection} \label{sec:LD}

Experimental upper limits on possible violations of local Lorentz
invariance are extremely tight (\cite{Wil93}), so that we are justified in 
assuming that photons follow 4D null geodesics, $ds=0$.  The situation 
is not so clear with regard to the 5D line element.  However, it is 
economical to follow KWE and suppose that {\em all particles 
follow ND null geodesics in N-dimensional gravity\/}, whether massive 
or not.\footnote[1]{This assumption is supported by various lines
   of argument.  In one version of 5D gravity, for example, the fifth 
   coordinate $y$ is related to {\em rest mass\/} $m$ (\cite{Wes84}), so 
   that one has $dS^{\, 2} = ds^2+g_{44}(G/c^2)^2 \, dm^2$.  If all 
   particles move on 5D null geodesics, then $ds^2=-g_{44}(G/c^2)^2 \, dm^2$.
   It then follows that $ds=0$ for photons, which have $m=$~const$=0$.
   For massive particles with $ds \neq 0$, one expects variations in 
   rest mass $m$, which are however below currently detectable levels,
   owing to the small size of the dimension-transposing constant $G/c^2$
   (Overduin and Wesson 1997, 1998).  Recent work on incorporating 
   non-relativistic {\em quantum theory\/} into higher-dimensional 
   gravity also strongly suggests that all test particles travel on 
   ND null geodesics in the classical limit (\cite{Sea00}).}
Proceeding on this assumption, and substituting $ds=dS=0$ into 
equation~(\ref{dS_ds}), we get $dy=0$ also, so that
$\ell,h \rightarrow \infty$ and $k$ is undefined.  The ratios $\ell/h$ 
and $k/h$ are however well-behaved, and read
\begin{eqnarray}
\frac{\ell}{h} \! \! \! \! & = & \! \! \! \! r^{-2}A^{2a+b-1} \, 
   \frac{dt}{d\phi } = \mbox{ finite} \; , \nonumber \\
\frac{k}{h} \! \! \! \! & = & \! \! \! \! r^{-2}A^{a+2b-1} \, 
   \frac{dy}{d\phi }=0 \; .
\end{eqnarray}
For self-consistency, therefore, the terms in $k/h$ can be dropped from 
KWE equations~(7),(8),(11) and (12).  Equation~(8) of that paper, 
in particular, reduces to
\begin{equation}
\left( \frac{du}{d\phi }\right) ^2+Au^2-\frac{\ell ^{\, 2}}{h^2} 
   A^{2-2a-b} = 0 \; ,
\label{NewKal8}
\end{equation}
and the definition of the parameter $p$, KWE equation~(12), becomes just
\begin{equation}
p \equiv -(2-2a-b) \, \frac{\ell ^{\, 2}}{h^2} \; .
\label{New_p}
\end{equation}
The photon's trajectory is deflected by an angle
\begin{equation}
\delta\phi = \omega = \frac{4M}{r_{\rm o}}+2Mpr_{\rm o} \; ,
\label{Kal18.1}
\end{equation}
which agrees with KWE equation~(18.1).
At the closest approach to the central body, we have
$u = u_o = 1/r_{\rm o}$ and $du/d\phi = 0$, so that
equation~(\ref{NewKal8}) gives
\begin{equation}
\frac{\ell ^{\, 2}}{h^2}=A_{\rm o}^{2a+b-1}u_o^2 = \frac 1{r_{\rm o}^2} +
   O ( \varepsilon ) \; ,
\label{ell-squared}
\end{equation}
where $\varepsilon\equiv M$ is a small parameter.  Putting 
equations~(\ref{New_p}) and (\ref{ell-squared}) into equation~(\ref{Kal18.1}),
we find for the final light deflection angle
\begin{equation}
\delta \phi = (4a+2b) \frac{M}{r_{\rm o}} + O (\varepsilon ^2) \; ,
\label{LightDef}
\end{equation}
as in KWE equation~(18.2), where 
however it is presented as a special case $k=0$.  We see here that 
equation~(\ref{LightDef}) is in fact entirely {\em general\/} for 
light deflection, and does not depend on any choice of $k$, which 
is in any case undefined when $ds=dS=0$.

To obtain experimental constraints from the light deflection result,
let us express equation~(\ref{LightDef}) for the Sun in terms of the
deviation $\Delta_{\scriptscriptstyle {\rm LD}}$ from the general relativity 
prediction $\delta\phi_{\scriptscriptstyle {\rm GR}}$, as follows
\begin{equation}
\delta\phi = \delta \phi_{\scriptscriptstyle {\rm GR}} ( 1 + 
   \Delta_{\scriptscriptstyle {\rm LD}} ) \; ,
\end{equation}
where (to first order in $\varepsilon$):
\begin{eqnarray}
\delta\phi_{\scriptscriptstyle {\rm GR}} \! \! \! \! & \equiv & 
   \! \! \! \! 4 M_{\scriptscriptstyle \odot}/r_{\rm o} \; , 
   \nonumber \\
\Delta_{\scriptscriptstyle {\rm LD}} \! \! \! \! & \equiv & \! \! \! \! 
   a + b/2 - 1 \; .
\label{Delta_LD}
\end{eqnarray}
Using the consistency relation~(\ref{Consist}) we find
\begin{equation}
a = - b/2 \pm (1 - 3b^2/4)^{1/2} \; .
\label{a_b_temp}
\end{equation}
Theoretical and numerical work indicates that $|b| \ll 1$ in the
solar system (\S \ref{sec:Disc}), and our experimental limits
bear this out.  The negative roots of equation~(\ref{a_b_temp})
may also be ignored, as they are inconsistent with the limiting
Schwarzschild case, and also imply the possibility of negative
gravitational and/or inertial soliton mass (\cite{Gro83,Lim95,Ove97}).
We therefore take
\begin{equation}
a = 1 - b/2 - 3b^2/8 + O(b^4) \; ,
\label{a_b}
\end{equation}
in the solar system, whereupon equation~(\ref{Delta_LD}) gives
\begin{equation}
\Delta_{\scriptscriptstyle {\rm LD}} = -3b^2/8 + O(b^4) \; .
\label{Delta_ld}
\end{equation}
The best available constraints on $\Delta_{\scriptscriptstyle {\rm LD}}$
come from long-baseline radio interferometry, which implies that
$|\Delta_{\scriptscriptstyle {\rm LD}}| \leq 0.0017$ (\cite{Rob91,Leb95}).
We therefore infer an upper limit
\begin{equation}
|b| \leq 0.07 \; ,
\end{equation}
for the Sun.  This could potentially be tightened by more than an order of 
magnitude using a proposed astrometric optical interferometer sensitive
to departures from Einstein's theory of as little as
$|\Delta_{\scriptscriptstyle {\rm LD}}| \leq 10^{-5}$ (\cite{Rea86}).

It is important to bear in mind, however, that the parameter $b$
characterizing the soliton metric~(\ref{Sol_Metric}) is not a universal
constant of nature like $G$ or $c$, but may in principle vary from soliton
to soliton.  Kaluza-Klein gravity as an alternative to 4D general relativity
is therefore best constrained by the application of two or more tests to 
the {\em same system\/}.
With this in mind we can use a recent measurement of light deflection by 
Jupiter, for which $|\Delta_{\scriptscriptstyle {\rm LD}}| \leq 0.17$ 
(\cite{Tre91}), to obtain
\begin{equation}
|b| \leq 0.7 \; ,
\end{equation}
for that planet.
It has also been proposed to measure light deflection by the {\em Earth\/}
using the Hipparcos satellite, with an estimated precision of 
12\% (\cite{Gou93}).  Such a test would be sensitive to values of 
$|b| \leq 0.6$ for the Earth.  The Gravity Probe~B satellite should
also be able to detect this effect by means of its guide star telescope,
though with a somewhat lower precision (\cite{Adl00}).

\section{Time delay} \label{sec:TD}

The arguments in the previous section regarding the parameter $k$ also apply
to the time delay (or radar ranging) test, and circular photon orbits as well.
That is, terms in $k/h$ and $k/\ell$ may be dropped from KWE equations~(20-24)
for radar ranging, and KWE equations~(28-30) for circular orbits.
The final results given there, however --- equations~(25) and (31) 
respectively --- are correct.  In fact, they hold not only for the special
case $k=0$, but quite generally.

In particular, the excess round-trip time delay $\Delta${\large $\tau$} for 
signals emitted from Earth (at distance $r_{\rm e}$ from the Sun) which 
graze the Sun (at nearest distance $r_{\rm o}$) and bounce off another 
planet (at $r_{\rm p}$) may be calculated by setting $k/\ell$=0 in KWE 
equation~(24) to obtain
\begin{equation}
\Delta \mbox{\large $\tau$} = \Delta 
   \mbox{\large $\tau$}_{\scriptscriptstyle {\rm GR}}
   (1 + \Delta_{\scriptscriptstyle {\rm TD}} ) \; , 
\label{TimeDelay}
\end{equation}
where (to first order in $\varepsilon$)
\begin{eqnarray}
\Delta \mbox{\large $\tau$}_{\scriptscriptstyle {\rm GR}} \! \! \! \! 
& \equiv & \! \! \! \! 4 M_{\scriptscriptstyle \odot} \left[ \ln \left( 
   \frac{r_{\rm p} + \sqrt{r_{\rm p}^2 - r_{\rm o}^2}}{r_{\rm o}} \right)
   \right. \nonumber \\
& & \hspace{6mm} + \ln \left. \left( \frac{r_{\rm e} + \sqrt{r_{\rm e}^2 - 
   r_{\rm o}^2}}{r_{\rm o}} \right) \right] \; , \nonumber \\
\Delta_{\scriptscriptstyle {\rm TD}} \! \! \! \! & \equiv & \! \! \! \! 
   a + b/2 -1 \nonumber \\
& = & \! \! \! \! -3b^2/8 + O(b^4) \; .
\end{eqnarray}
We note that departures from 4D general relativity for time delay have 
exactly the same form as they do for light deflection.

The best experimental constraint on time delay so far has come from the 
Viking lander on Mars, and gives 
$|\Delta_{\scriptscriptstyle {\rm TD}}| \leq 0.002$ (\cite{Rea79}).
This leads immediately to the upper bound
\begin{equation}
|b| \leq 0.07 \; ,
\end{equation}
for the Sun, exactly the same as the limit obtained in the case of light 
deflection using long-baseline interferometry.

Keeping in mind that values of $b$ can differ from soliton to soliton,
however, it is possible that different physical setups could provide new
information.  For instance, one could attempt to measure $b$ for the Earth
by sending grazing signals from an orbiting satellite past our planet and
bouncing them off the Moon; retroflectors left there by Apollo astronauts 
are routinely used for lunar laser ranging (\cite{Wil96}).  
Substituting $M_{\rm e}$ for $M_{\scriptscriptstyle \odot}$ and replacing
$r_{\rm e}, r_{\rm p}$ and $r_{\rm o}$ with the appropriate distances, 
we find an expected excess time delay of order 400~ps using a satellite in
geostationary orbit.  This is well above the currently available resolution
of $\sim 50$~ps (\cite{Sam98}).  The feasibility of such a proposal would
likely be limited by the weakness of the reflected signal.  Better results
might be obtained by active ranging between two orbiting satellites, or by
statistical analysis of ranging data between two such satellites and an
Earth station (the latter would however require excellent atmospheric
modelling).

In the same vein, one might attempt to measure $b$ for the Moon
by grazing it with signals from the Earth and bouncing them off the 
Viking lander on Mars.  This might be done when Mars is at nearest 
approach (on the same side of the Sun as the Earth) to minimize signal
contamination from the competing effect of the Sun.  Substituting 
$M_{\rm m}$ for $M_{\scriptscriptstyle \odot}$ in equation~(\ref{TimeDelay}),
however, and replacing $r_{\rm e}, r_{\rm p}$ and $r_{\rm o}$ with the 
appropriate distances, we find that the Moon's excess time delay
(of order 10~ps) would be so short as to make this a daunting
task at present.

\section{Perihelion advance} \label{sec:PP}

We now switch our attention to massive test particles.  In terms of 
a new variable $u\equiv 1/r$, equation~(\ref{r-DE}) becomes
\begin{eqnarray}
\left( \frac{du}{d\phi }\right) ^2 \! \! \! \! \! \! \! & + & \! \! \! \!
   A u^2 - \left( \frac{\ell ^{\, 2}}{h^2} A^{2-2a-b} - \frac{k^2}{h^2} 
   A^{2-a-2b} \right. \nonumber \\
& - & \! \! \! \! \left. \frac{1}{h^2} A^{2-a-b}
   \right) = 0 \; .
\label{peri-DE}
\end{eqnarray}
Differentiating with respect to $\phi$ (and letting primes denote $d/d\phi$), 
we find that noncircular orbits ($u^{\prime}\neq 0$) are governed by the 
following differential equation
\begin{equation}
u^{\prime \prime } + (1+\gamma \epsilon ) \, u = B + \epsilon B^{-1} u^2 + 
   O\left( \epsilon ^2\right) \; ,
\label{u-DE}
\end{equation}
where five new quantities have been introduced
\begin{eqnarray}
\gamma \! \! \! \! & \equiv & \! \! \! \! -\frac{f}{3d} \; , \; \; \; 
   \epsilon \equiv 3 \, M B \; , \; \; \; 
B \equiv \frac{Md}{h^2} \, , \nonumber \\
d \! \! \! \! & \equiv & \! \! \! \! (2-a-b) -\ell ^{\, 2} (2-2a-b) 
   \nonumber \\
   & & \! \! \! \! + k^2 (2-a-2b) \; , \nonumber \\
f \! \! \! \! & \equiv & \! \! \! \! 2 \, (2-a-b)(-1+a+b) \nonumber \\
   & & \! \! \! \! + \, 2\ell ^{\, 2}(-2+2a+b)(-1+2a+b) \nonumber \\
   & & \! \! \! \! + \, 2k^2 (2-a-2b)(-1+a+2b) \; .
\label{gamma_B_d_f}
\end{eqnarray}
These expressions agree with KWE equations~(32-36).
(We have however chosen to relabel their $e$ as $f$,
for reasons that will become clear shortly.)

The solution of the differential equation~(\ref{u-DE}) is
\begin{eqnarray}
u = \frac{1}{r} = B \! \! \! \! & + & \! \! \! \! \left( 1-\frac{\gamma}{2} 
   \right) C \cos \left\{ \left[ 1-\epsilon \, \left( 1-\frac{\gamma}{2} 
   \right) \right] \phi \right\} \nonumber \\
& + & \! \! \! \! \epsilon \, (1-\gamma ) B + \epsilon \, 
   \frac{C^{\, 2}}{2B} \left( 1-\frac{\gamma}{2} \right) ^2 \nonumber \\ 
& - & \! \! \! \! \epsilon \, \frac{C^{\, 2}}{6B}\left( 1- 
   \frac{\gamma}{2} \right) ^2 \! \! \cos 2\phi + O( \epsilon ^2) \; ,
\label{u-sol'n}
\end{eqnarray}
where $C$ is an integration constant.  [This result differs slightly from 
KWE equation~(37), where the factors of $(1-\gamma/2)^2$ were omitted.] 
Equation~(\ref{u-sol'n}) can be written in a physically more transparent 
form by introducing two new quantities $e$ and $\omega$ via
\begin{equation}
\left( 1-\frac{\gamma}{2}\right) C\equiv B \, e \; , \; \; \; 
   1-\epsilon \left( 1-\frac{\gamma}{2}\right) \equiv \omega \; .
\end{equation}
With these definitions, we find that
\begin{eqnarray}
u = \frac{1}{r} = B (1 \! \! \! \! & + & \! \! \! \! e \cos\omega\phi ) +
   \frac{1}{2} \varepsilon B^2 \! \left[ -e^2 \cos 2\omega \phi 
   \right. \nonumber \\
& + & \! \! \! \! \left. 6 \left( 1 - \gamma + \frac{1}{2} 
   e^2 \right) \right] + O(\varepsilon ^2) \; ,
\label{u-sol'n2}
\end{eqnarray}
where $\varepsilon\equiv M$ is a small parameter as before, and
\begin{equation}
\omega = 1 - 3\varepsilon B \left( 1 + \frac{f}{6d} \right) + 
   O(\varepsilon ^2) \; .
\end{equation}
The first term on the right-hand side of equation~(\ref{u-sol'n2})
is of order $\varepsilon^0$, and shows explicitly the elliptical shape
of the orbit.  This is then modified by the second term, of order 
$\varepsilon^1$.  Note that $e$ is just the {\em eccentricity\/} of the 
ellipse.  The angular shift between two successive perihelia is given by
\begin{equation}
\delta \phi = \phi -2\pi = 
   \frac{6\pi M^2d}{h^2}\left( 1+\frac{f}{6d}\right) +
   O(\varepsilon ^2) \; ,
\label{peri_shift}
\end{equation}
in agreement with the final result~(38.1) of KWE.  It should be
emphasized that the angular momentum $h$ is not in general the same quantity
in 5D as it is in 4D.  In particular, putting equations~(\ref{ddS_dds})
and (\ref{Sol_Metric}) into the second of equations~(\ref{ell_h_k}),
we find
\begin{eqnarray}
h \! \! \! \! & = & \! \! \! \! A^{1-a-b}r^2\frac{d\phi }{dS} = 
   A^{1-a-b}r^2\frac{d\phi }{ds} \sqrt{ 1+A^{-b}k^2} \nonumber \\
& = & \! \! \! \! h_{\scriptscriptstyle {\rm (4D)}}
   \sqrt{1+A^{-b}k^2} \; .
\end{eqnarray}
If $k\neq 0$, therefore, it follows that 
$h\neq h_{\scriptscriptstyle {\rm (4D)}}$.

To eliminate $h$ from equation~(\ref{peri_shift}), let us consider
the points along the orbit where $r$ takes its minimum value $r_-$ 
and maximum value $r_+$ respectively.  From inspection of 
equation~(\ref{u-sol'n2}) we see that
$r_- = B^{-1}(1+e)^{-1} + O(\varepsilon )$ at $\omega\phi=0$ and
$r_+ = B^{-1}(1-e)^{-1} + O(\varepsilon )$ at $\omega\phi=\pi$.
The semimajor axis $a_{\rm o}$ of the ellipse is then
\begin{equation}
a_{\rm o} \equiv \frac{1}{2} \, (r_{-}+r_{+}) = \frac{1}{B(1-e^2)} + 
   O(\varepsilon ) \; ,
\end{equation}
so that
\begin{equation}
B \equiv \frac{Md}{h^2} = \frac{1}{a_{\rm o} (1-e^2)} + O(\varepsilon ) \; ,
\label{h_to_ao}
\end{equation}
or
\begin{equation}
h^2 = \varepsilon (1-e^2) \, a_{\rm o} \, d + O(\varepsilon ^2) \; .
\label{h_vareps}
\end{equation}
Substituting equation~(\ref{h_to_ao}) into equation~(\ref{peri_shift}),
we find
\begin{equation}
\delta \phi = \frac{6\pi M}{a_{\rm o}(1-e^2)} \left( 1+\frac{f}{6d}\right) +
   O(\varepsilon ^2) \; .
\label{Final_precession}
\end{equation}
Only one term in this result remains physically obscure, and that is the 
ratio $f/d$.  This is given in terms of $\ell$ and $k$ by the 
definitions~(\ref{gamma_B_d_f}).  The latter two constants are
related by equation~(\ref{peri-DE}) as follows
\begin{equation}
\ell ^{\, 2} = h^2\left[ \left( \frac{du}{d\phi }\right) ^2+Au^2\right]
   A^{2a+b-2}+k^2A^{a-b}+A^a \; .
\label{PhysObscure}
\end{equation}
Since $h^2$ is of order $\varepsilon^1$ by equation~(\ref{h_vareps}), 
while $u$ and $u^{\prime }$ are of order $\varepsilon^0$ by 
equation~(\ref{u-sol'n2}), it follows from equation~(\ref{PhysObscure}) that
$\ell ^{\, 2} = 1+k^2 + O(\varepsilon )$.
Using the definitions~(\ref{gamma_B_d_f}), we therefore obtain
\begin{equation}
\frac{f}{6d} = -1 + a + \frac{2b}{3} + \frac{k^2b\left( a-b\right) /3}
   {a+k^2\left( a-b\right) } + O(\varepsilon ) \; ,
\end{equation}
so that the final perihelion precession angle~(\ref{Final_precession}) 
becomes
\begin{equation}
\delta \phi = \frac{6\pi M}{a_{\rm o} (1-e^2)} \left[ a + \frac{2}{3} \, b + 
   \frac{k^2 (a-b) b/3}{a+k^2 (a-b) }\right] +
   O(\varepsilon ^2) \; .
\end{equation}
This represents the generalization of KWE equation~(38.2)
to cases in which $k \neq 0$ (and eccentricity $e \neq 0$).  In the special
case $b=0$ (and $a=+1$), for which the metric~(\ref{Sol_Metric}) reduces 
to Schwarzschild form on hypersurfaces $y=$~const, it is interesting to
note that one recovers the standard 4D general relativity result, regardless
of the value of $k$.  In this limit, therefore, the perihelion shift test 
is insensitive to the momentum of the test body along the extra coordinate.
And in general, one must choose a soliton with $b \neq 0$ in order to 
distinguish experimentally between test particles with different values 
of $k$.

As usual, let us parametrize our result in terms of the departure from
4D general relativity so that
\begin{equation}
\delta \phi = \delta \phi_{\scriptscriptstyle {\rm GR}}
   (1 + \Delta_{\scriptscriptstyle {\rm PP}} ) \; , 
\label{PeriShift}
\end{equation}
where (to first order in $\varepsilon$):
\begin{eqnarray}
\delta\phi_{\scriptscriptstyle {\rm GR}} \! \! \! \! & \equiv & \! \! \! \!
   \frac{6\pi M} {a_{\rm o} (1-e^2)} \; , \nonumber \\
\Delta_{\scriptscriptstyle {\rm PP}} \! \! \! \! & \equiv & \! \! \! \! a + 
   \frac{2}{3} \, b + \frac{k^2 (a-b) b/3}{a + k^2 (a-b)} - 1 \; .
\label{Delta_pp}
\end{eqnarray}
Theoretical work indicates that $k$, which is a measure of momentum along 
the fifth dimension, is related to the {\em charge-to-mass ratio\/} of the 
test body (\cite{Wes97}).  For a planet such as Mercury, we may take $k=0$.
Putting equation~(\ref{a_b}) into equation~(\ref{Delta_pp}), we therefore have
\begin{equation}
\Delta_{\scriptscriptstyle {\rm PP}} = b/6 - 3b^2/8 + O(b^4) \; .
\end{equation}
Perihelion precession is thus a potentially more sensitive probe of
higher-dimensional gravity than either light deflection or time delay,
in that it depends on the {\em first\/}, as well second order in $b$.

Unfortunately, however, this increased sensitivity is offset in the case 
of Mercury's orbit about the Sun by uncertainty in the solar oblateness.
The latter introduces a new term $\xi J_2$ inside the brackets on the 
right-hand side of equation~(\ref{PeriShift}), where
$\xi \equiv R_{\odot}^2/2 M_{\odot} a_{\rm o} (1-e^2)$
and $J_2$ is the solar quadrupole moment (\cite{Cam83}).
Dividing through by the orbital period $T$, we may therefore write 
for the rate of perihelion advance (to order $b^{\, 3}$)
\begin{equation}
\Delta \omega \equiv \frac{\delta \phi}{T} = 
   \Delta \omega_{\scriptscriptstyle {\rm GR}}
   (1 + \xi \, J_2 + b/6 - 3b^2/8 ) \; , 
\end{equation}
where $\Delta\omega_{\scriptscriptstyle {\rm GR}} \equiv 
\delta\phi_{\scriptscriptstyle {\rm GR}}/T=42.98$~arcsec/century.
The observed value of Mercury's perihelion precession rate is quite
close to this value, $\Delta\omega = 43.11 \pm 0.21$~arcsec per century
(Shapiro, Counselman and King 1976).
Experimental data on $J_2$ is a good deal more controversial
and has ranged over two orders of magnitude, from a maximum value of
$(23.7 \pm 2.3) \times 10^{-6}$ (\cite{Dic67}) to a minimum of
$(0.17 \pm 0.02) \times 10^{-6}$ (\cite{Duv84}).  One straightforward
least-sqares fit to a number of published measurements leads to
intermediate value of $J_2=5.0 \times 10^{-6}$, which however implies
a general relativistic precession rate more than two standard deviations 
away from that observed (\cite{Cam83}).  Such a discrepancy could be explained
in the context of higher-dimensional gravity by modelling the Sun as a 
soliton with $b=-0.062$.  This is just consistent with the constraint 
$|b| \leq 0.07$ from light deflection (\S \ref{sec:LD}) and time delay
(\S \ref{sec:TD}), which is intriguing since these tests probe somewhat 
independent aspects of relativistic gravity.  Improved experimental data
relating to any of the three tests would be of great interest.

Conservative limits on $b$ from perihelion precession may be obtained
by quoting the results of a recent review in which all available data
(to 1997) have been combined to give a weighted mean value for the solar
oblateness of $J_2 = (3.64 \pm 2.84) \times 10^{-6}$ (\cite{Roz97}).
Using this uncertainty range, together with that in the observed value 
of $\Delta\omega$ for Mercury's orbit, we find that
\begin{equation}
b = -0.03 \pm 0.07 \; ,
\end{equation}
for the Sun.  This is consistent with the bounds obtained from light
deflection and time delay.  Sensitivity of the perihelion precession test
to the value of $b$ could be improved by an order of magnitude if better
data on $J_2$ were to become available; the proposed ASTROD mission, 
for example, might measure this parameter to an accuracy of 
$5 \times 10^{-8}$ (\cite{Ni98}).

\section{Geodetic effect} \label{sec:GE}

We now move on to consider {\em spinning\/} massive test particles with 
velocity 5-vectors $u^C\equiv dx^C/dS$ and spin 5-vectors $S^{\, C}$.
The motion of these objects is governed by three central equations; 
namely, the geodesic equation
\begin{equation}
\frac{d^2x^C}{dS^{\, 2}} + \widehat{\Gamma }_{AB}^C \, u^A \, u^B = 0 \; ,
\label{Geodesic}
\end{equation}
the parallel transport equation
\begin{equation}
\frac{dS^{\, C}}{dS} + \widehat{\Gamma }_{AB}^C \, S^A\, u^B = 0 \; ,
\label{ParaTrans}
\end{equation}
and the orthogonality condition
\begin{equation}
u^C \, S_C = 0 \; .
\label{Spin}
\end{equation}
Here $\widehat{\Gamma }_{AB}^C$ refers to the 5D Christoffel symbol
for the metric~(\ref{Sol_Metric}).  This is defined in exactly the same
manner as the usual 4D Christoffel symbol, with indices running over
five values instead of four (see KWE, Appendix~A1 for
details\footnote[2]{There are some minor
   typographical errors in this appendix, which we note briefly here.
   The factors of $(1-2M)/r$ in equations~(A2.2), (A2.6) and (A2.7) 
   should read $1-2M/r$.  The same thing applies to equations~(57)
   and (58) in the main body of KWE.  Also, the exponents
   $-(1/2)$ and $1/2$ in equations~(57) and (58) should be switched,
   in agreement with equations~(A2.7) and (A2.2) respectively.
   These discrepancies do not affect any of the other equations or conclusions
   reported in KWE, and do not appear in the new reference book on 
   Kaluza-Klein gravity by Wesson (1999).}).

In order to simplify the problem, we follow KWE in assuming 
that the test particle moves in a circular orbit with $\theta =\pi /2$,
$r = r_{\rm o}$ and $\dot{\theta }=\dot{r}=0$.  Its velocity $u^C$ may then 
be expressed as follows in terms of the constants of motion $\ell,h$ 
and $k$, as given by equations~(\ref{ell_h_k})
\begin{equation}
u^C \equiv \frac{dx^C}{dS} = \left( \ell A^{-a}, \; 0, \; 0, \; 
   hr_{\rm o}^{-2}A^{a+b-1}, \; kA^{-b} \right) \; .
\label{Velocity}
\end{equation}
From the metric~(\ref{Sol_Metric}), we have
\begin{equation}
1 = A^a\left( u^0\right) ^2 - A^{1-a-b}r_{\rm o}^2\left( u^3\right) ^2 - 
   A^b\left( u^4\right) ^2 \; ,
\end{equation}
which, with equation~(\ref{Velocity}), implies
\begin{equation}
\ell ^{\, 2}-h^2r_{\rm o}^{-2} A^{2a+b-1} - k^2 A^{a-b}-A^a = 0 \; .
\label{Satisfactory}
\end{equation}
It may be shown that the motion of the test body as given by 
equations~(\ref{Velocity}) and (\ref{Satisfactory}) is geodesic in the
sense of equation~(\ref{Geodesic}).

We now propose to generalize the treatment of KWE by leaving the extra
component $S^{\, 4}$ of spin unrestricted, rather than setting it to zero.
In fact, writing explicitly $S^{\, C} \equiv (S^{\, 0},S^{\, 1},S^{\, 2},
S^{\, 3},S^{\, 4})$, we find that the orthogonality condition~(\ref{Spin})
imposes the following restriction on the spin components
\begin{equation}
\ell \, S^{\, 0} - h \, S^{\, 3} - k \, S^{\, 4} = 0 \; ,
\label{Spin_Consist}
\end{equation}
so that $S^{\, 4}$ will not vanish in general, if the parameter
$k$ is well-defined.

We now proceed to solve the parallel transport equation~(\ref{ParaTrans}),
taking one value of the index~$C$ at a time.
To begin with, the $C=2$ component gives
\begin{equation}
S^{\, 2} = \frac{H_2}{r_{\rm o}} =\mbox{ const} \; , \; \; \;
   H_2 = \mbox{ const} \; .
\label{C=2_comp}
\end{equation}
(Note that, due to our choice of coordinates, $S^{\, 0},S^{\, 1}$ and
$S^{\, 4}$ are dimensionless while $S^{\, 2}$ and $S^{\, 3}$ have units
of inverse length.)
Defining a new function $g=g(S)$ of the 5D proper time, we may write
without loss of generality
\begin{equation}
S^{\, 0} \equiv H_0 \, g \, , \; \; \; 
   H_0 = \mbox{ const} \; .
\label{C=0_comp}
\end{equation}
The $C=0$ component of equation~(\ref{ParaTrans}) then reads
\begin{equation}
S^{\, 1} = -\frac{H_0}{a \ell M} \, r_{\rm o}^2 \, A^{a+1} \, \frac{dg}{dS} \; .
\label{C=1_comp}
\end{equation}
The $C=4$ component, meanwhile, takes the form
\begin{eqnarray}
S^{\, 4} \! \! \! \! & = & \! \! \! \! H_4 \, g + K_4 \; , \nonumber \\
H_4 \! \! \! \! & = & \! \! \! \! \frac{bk}{a \ell} \, H_0 \, A^{a-b} 
   = \mbox{ const} \; , \nonumber \\
K_4 \! \! \! \! & = & \! \! \! \! \mbox{ const} \; ,
\label{C=4_comp}
\end{eqnarray}
where we have used equation~(\ref{C=1_comp}).
In a similar way, the $C=3$ component of equation~(\ref{ParaTrans}) gives
\begin{eqnarray}
S^{\, 3} \! \! \! \! & = & \! \! \! \! \frac{H_3}{r_{\rm o}} g + 
   K_3 \; , \nonumber \\
H_3 \! \! \! \! & = & \! \! \! \! \frac{hH_0}{a \ell M} \, A^{2a+b-1} 
   \left[ 1 - (1+a+b) \, \frac{M}{r_{\rm o}} \right] = \mbox{ const} \; , 
   \nonumber \\
K_3 \! \! \! \! & = & \! \! \! \! \mbox{ const} \; .
\label{C=3_comp}
\end{eqnarray}

We now solve the $C=1$ component of equation~(\ref{ParaTrans}), assuming for 
simplicity that $K_3 = K_4 = 0$.  Using equations~(\ref{Velocity}),
(\ref{C=0_comp}), (\ref{C=4_comp}) and (\ref{C=3_comp}), we find
\begin{eqnarray}
\frac{dS^{\, 1}}{dS} \! \! \! \! & = & \! \! \! \! - \frac{H_0M}{a \ell 
   \, r_{\rm o}^2} \, A^{a+b-1} \left\{ a^2 \ell ^{\, 2} - b^2 k^2 A^{a-b} 
   \frac{ }{ } \right. \nonumber \\
& - & \! \! \! \! \! \left. \frac{h^2}{M^2} \! 
   \left[ 1 - (1+a+b) \, \frac{M}{r_{\rm o}} \right] ^2 \! \! \! \! 
   A^{2a+b-1} \! \right\} \! g \, .
\end{eqnarray}
Differentiating equation~(\ref{C=1_comp}) with respect to $S$, meanwhile, gives
\begin{equation}
\frac{dS^{\, 1}}{dS} = -\frac{H_0}{a \ell M} \, r_{\rm o}^2 \, A^{a+1} \,
   \frac{d^2g}{dS^{\, 2}} \; .
\end{equation}
Equating these two expressions, we obtain
\begin{equation}
\frac{d^2g}{dS^{\, 2}} = -\Omega ^2 \, g \; ,
\label{g-DE}
\end{equation}
where
\begin{eqnarray}
\Omega ^2 \! \! \! & \equiv & \! \! \! \! \frac{h^2}{r_{\rm o}^4} \, 
   A^{b-2} \left\{ \left[ 1 - (1+a+b) \, \frac{M}{r_{\rm o}} \right] ^2 
   \! \! \! A^{2a+b-1} \right. \nonumber \\
& & \! \! \! \! \left. - \frac{M^2}{h^2} \left( a^2 \ell ^{\, 2} - b^2 \, k^2 \, 
   A^{a-b} \right) \right\} \; .
\label{Om_squared}
\end{eqnarray}
The general solution of equation~(\ref{g-DE}) is $g(S)=K_1 \sin (\Omega S)
+ K_2 \cos(\Omega S)$.  We choose $K_1=1$ and $K_2=0$ for simplicity.
The spin components are then given by
\begin{eqnarray}
S^{\, 0} \! \! \! \! & = & \! \! \! \! H_0 \sin (\Omega S) \; , \; \; \;
   S^{\, 1} = H_1 \cos (\Omega S) \; , \nonumber \\
S^{\, 2} \! \! \! \! & = & \! \! \! \! \frac{H_2}{r_{\rm o}} \; , \; \; \;
   S^{\, 3} = \frac{H_3}{r_{\rm o}} \sin (\Omega S) \; , \nonumber \\
S^{\, 4} \! \! \! \! & = & \! \! \! \! H_4 \sin (\Omega S) \; ,
\end{eqnarray}
where $H_1$ and $H_2$ are arbitrary constants and
\begin{eqnarray}
H_0 \! \! \! \! & = & \! \! \! \! -\frac{a \ell M}{r_{\rm o}^2\Omega } \, 
   A^{-a-1} \, H_1 \; , \nonumber \\
H_3 \! \! \! \! & = & \! \! \! \! -\frac{h}{r_{\rm o}^2\Omega } \, A^{a+b-2} 
   \left[ 1 - (1+a+b) \, \frac{M}{r_{\rm o}} \right] H_1 \; , \nonumber \\
H_4 \! \! \! \! & = & \! \! \! \! -\frac{bkM}{r_{\rm o}^2\Omega } \, 
   A^{-b-1} \, H_1 \; .
\label{3_Hs}
\end{eqnarray}
The spatial part of $S^{\, C}$ is thus seen to rotate in the plane of
the orbit with angular speed $\Omega$.
Substituting these results into equation~(\ref{Spin_Consist}) yields
\begin{eqnarray}
a \ell ^{\, 2} \! \! \! \! & - & \! \! \! \! \frac{h^2}{M r_{\rm o}} \, 
   A^{2a+b-1} \left[ 1 - (1+a+b) \, \frac{M}{r_{\rm o}} \right] 
   \nonumber \\
& - & \! \! \! \! b k^2 A^{a-b} = 0 \; .
\end{eqnarray}
Solving simultaneously with equation~(\ref{Satisfactory}),
we obtain for the constants of motion
\begin{eqnarray}
\ell ^{\, 2} \! \! \! \! & = & \! \! \! \! A^a \! \left\{ \! 1 + k^2 A^{-b} + 
   \frac{M}{r_{\rm o}} \! \left[ \! \frac{a+(a-b) k^2 A^{-b}}{1 - (1+2a+b) 
   M/r_{\rm o}} \right] \! \right\} \! , \nonumber \\
h^2 \! \! \! \! & = & \! \! \! \! M r_{\rm o} \, A^{1-a-b} \, \left[ 
   \frac{a+(a-b) \, k^2 A^{-b}} {1-\left( 1+2a+b \right) M/r_{\rm o}} 
   \right] \; .
\label{h_ell}
\end{eqnarray}
These expressions can be written in terms of a small parameter 
$\varepsilon \equiv M$ as usual
\begin{eqnarray}
\ell ^{\, 2} \! \! \! \! & = & \! \! \! \! (1+k^2) \left[ 1 - \left( a - 
   \frac{bk^2}{1+k^2} \right) \frac{M}{r_{\rm o}} \right] + 
   O(\varepsilon ^2) \; , \nonumber \\
h^2 \! \! \! \! & = & \! \! \! \! M r_{\rm o} \left[ a+(a-b) \, k^2 \right] 
   \left\{ 1 + \left[ (4a+3b-1) \right. \right. \nonumber \\ 
& & \! \! \! \! \left. \left. + \frac{2b(a-b) \, k^2}{a+(a-b) \, k^2} 
   \right] \frac{M}{r_{\rm o}} \right\} + O(\varepsilon ^3) \; .
\label{h_ell_2}
\end{eqnarray}
With the aid of equation~(\ref{Om_squared}), we then find for the
angular speed of the spin vector
\begin{eqnarray}
\Omega \! \! \! \! & = & \! \! \! \! \sqrt{\frac{[ a+(a-b) \, k^2] M}
   {r_{\rm o}^3}} \left\{ 1 + \frac{M}{2r_{\rm o}} \left[ 3 \, (1-a-b) 
   \right. \right.  \nonumber \\
& & \! \! \! \! \left. \left. + \frac{b(a-b) \, k^2}{a+(a-b) \, k^2}
   \right] + O(\varepsilon ^2) \right\} \; .
\label{Omega=}
\end{eqnarray}
This quantity is not the same as the test body's {\em orbital\/}
angular speed, which is given in terms of the 5D proper time $dS$ as
\begin{eqnarray}
\omega \! \! \! \! & \equiv & \! \! \! \! 
   \frac{d\phi }{dS} = hr_{\rm o}^{-2} A^{a+b-1} \\
& = & \! \! \! \! \sqrt{ \frac{M}{r_{\rm o}^3} } \, A^{(a+b-1)/2}
   \sqrt{ \frac{a+(a-b) \, k^2 A^{-b}} {1-(1+2a+b) M/r_{\rm o}} } \; , 
   \nonumber
\end{eqnarray}
where we have used equations~(\ref{ell_h_k}) and (\ref{h_ell}). 
In terms of $\varepsilon$
\begin{eqnarray}
\omega \! \! \! \! & = & \! \! \! \! \sqrt{\frac{\left[ a+(a-b) \, k^2 
   \right] M} {r_{\rm o}^3}} \left\{ \left[ 1 + \frac{M}{r_{\rm o}} 
   \left( \frac{3-b}{2} \right. \right. \right. \nonumber \\
& & \! \! \! \! \left. \left. \left. + \frac{b(a-b) \, k^2}{a+(a-b) \, 
   k^2} \right) \right] + O(\varepsilon ^2) \right\} \; .
\label{omega=}
\end{eqnarray}
It is precisely the excess of $\Omega$ over $\omega$ that gives rise to the
geodetic effect.

Suppose the spin vector $S^{\, C}$ is initially oriented in the radial 
direction; ie, $H_2=0$ at $S=0$.
During one orbit, the test body's angular displacement $\phi$
goes from $0$ to $2\pi $, so that $\delta S = 2\pi / \omega$.
In the same period, $S^{\, 3}$ goes from its initial value of zero
at $S=0$ to its final value at proper time $S$.  To first order in 
$\varepsilon $, the spin vector has advanced through an angle 
\begin{eqnarray}
\delta\phi \! \! \! \! & = & \! \! \! \! \frac{r_{\rm o}[S^{\, 3}(S) - 
   S^{\, 3}(0)]}{S^{\, 1}(0)} + O(\varepsilon ^2) \; , \nonumber \\
& = & \! \! \! \! 2\pi\frac{H_3}{H_1} \left( \frac{\Omega}{\omega} - 1 
   \right) + O(\varepsilon ^2) \; , \nonumber \\
& = & \! \! \! \! -2\pi\left( \frac{\Omega}{\omega} - 1 \right) + 
   O(\varepsilon ^2) \; ,
\end{eqnarray}
where we have used equations~(\ref{3_Hs}), (\ref{h_ell_2}) and
(\ref{Omega=}).  Combining equations~(\ref{Omega=}) and (\ref{omega=}), 
we find that
\begin{equation}
\frac{\Omega}{\omega} - 1 = -\frac{3M}{2r_{\rm o}} \left[ a + \frac{2}{3} b +
   \frac{k^2 b(a-b)/3}{a+(a-b) \, k^2} \right] + O(\varepsilon ^2) \; ,
\end{equation}
so that the geodetic precession angle can finally be expressed as follows in 
terms of its deviation from the prediction of 4D general relativity
\begin{equation}
\delta\phi = \delta\phi_{\scriptscriptstyle {\rm GR}} \,
   (1 + \Delta_{\scriptscriptstyle {\rm GP}} ) \; ,
\label{Geo_Pred}
\end{equation}
where (to first order in $\varepsilon$)
\begin{eqnarray}
\delta\phi_{\scriptscriptstyle {\rm GR}} \! \! \! \! & \equiv & \! \! \! \! 
   3\pi M/r_{\rm o} \; , \nonumber \\
\Delta_{\scriptscriptstyle {\rm GP}} \! \! \! \! & \equiv & \! \! \! \! a + 
   \frac{2}{3} \, b + \frac{k^2 (a-b) b/3}{a + k^2 (a-b)} - 1 \; .
\end{eqnarray}
This represents the generalization of KWE equation~(66)
to cases in which $S^{\, 4} \neq 0$.  Deviations from 4D general 
relativity have exactly the same form for geodetic precession as they do
for perihelion precession.  Taking $k=0$ and using equation~(\ref{a_b}), 
as in \S \ref{sec:PP}, we find that
\begin{equation}
\Delta_{\scriptscriptstyle {\rm GP}} = b/6 - 3b^2/8 + O(b^4) \; .
\label{Delta_gp}
\end{equation}
Like the perihelion shift, geodetic precession depends on $b$ to first 
as well as second order, and is thus a potentially more sensitive probe 
of the theory than either light deflection or time delay.

The Gravity Probe~B satellite, currently scheduled for launch in early
2001, has been designed to measure deviations from 4D general relativity 
with a precision of better than 
$|\Delta_{\scriptscriptstyle {\rm GP}}| \leq 2.5 \times 10^{-5}$ 
(\cite{Buc96}). Using equation~(\ref{Delta_gp}), we find that this
corresponds to a sensitivity to values as small as
\begin{equation}
|b| \leq 1 \times 10^{-4} \; ,
\end{equation}
or better for the Earth --- a constraint some five hundred times stronger
than any other solar system bound obtained to date, and five {\em thousand\/}
times stronger than the only other Earth-based test (light deflection
using Hipparcos; \S \ref{sec:LD}).

We conclude this section by noting that a complementary analysis of
geodetic precession has been carried out for a static, 
spherically-symmetric 5D metric different from that given by 
equation~(\ref{Sol_Metric}), one in which the fifth dimension
is flat (Mashhoon, Liu and Wesson 1994; Mashhoon, Wesson and Liu 1998).
The inclusion of spin is of special importance in this case since the 
classical tests (based on the equations of motion) alone cannot 
discriminate between 4D and 5D effects.
The geodetic precession rate has been computed, and differs from the
4D Einstein value in the weak-field, low velocity limit (\cite{Liu96}).
A preliminary interpretation of the discrepancy indicates, however, that
it is likely to be somewhat below the threshold of detection by Gravity
Probe~B (\cite{Ove98}).

\section{Discussion} \label{sec:Disc}

Having obtained upper limits on $|b|$ of order $0.07$
(and possibly $10^{-4}$) from experiment, we consider here the 
range of values that might be expected for this parameter
on theoretical grounds.  These turn out to be small
(perhaps of order $10^{-8}$ to $10^{-2}$) in the solar system,
but could be larger (of order $0.1$) in larger
systems such as clusters of galaxies.

These estimates are based on the fact that the soliton's effective
4D mass is not concentrated at a point, like that of a black hole,
but has instead a finite (though sharply peaked) density profile 
whose steepness depends on the metric parameters (\cite{Liu92,Wes94}). 
Quoting the latter authors, but replacing their metric parameters
$\tilde{a},\epsilon,k$ (due to Davidson and Owen 1985) with our $M,a,b$
via $M \equiv 2/\tilde{a}$, $a \equiv \epsilon k$ and $b \equiv -\epsilon$,
we find for the density of the soliton
\begin{equation}
8\pi \rho (r) = \frac{-ab \, M^2/r^4}{\left[ 1-(M/2r)^2 \right]^4}
                \left( \frac{1-M/2r}{1+M/2r} \right)^{\! 2(a+b)} \; .
\label{Sol_Density}
\end{equation}
Pressure is given by $p=\rho/3$, so that the matter described by
equation~(\ref{Sol_Density}) could be radiationlike, or composed of 
ultrarelativistic particles such as neutrinos.  Total gravitational mass
(as deduced from the asymptotic form of the metric) is $M_g=aM$, 
so it is clear that $b$ must be {\em negative\/} for positive density.
Numerical analysis further reveals that the mass of the soliton is
increasingly concentrated at small $r$ as $|b|$ approaches zero,
and that the 4D Schwarzschild limit ($b=0$) can in fact
be viewed as a maximally compressed soliton (\cite{Wes94}).
Physically, this means that solar system bodies, which (viewed as
solitons) are essentially point masses, are likely to be associated
with very small values of $|b|$.

To attach some numbers to these qualitative remarks,
we make use of equation~(\ref{a_b}) and consider the weak-field 
($r \gg M/2$), small-$b$ limit, in which 
\begin{equation}
\rho (r) \approx -b \, G M_g^2 / (8\pi c^2 r^4) \; ,
\label{Sol_Dens2}
\end{equation}
where we have reverted to physical units.
Equation~(\ref{Sol_Dens2}) allows us to associate 
ranges of $b$-values with solitons of mass $M_g$, if the density $\rho$
can be estimated at some radius $r$.
It has, for instance, been suggested (eg, \cite{Fre86,Gou92}) that 
relativistic hot dark matter in the form of massive neutrinos could 
be trapped inside the Earth.  Krauss et al. (1986) have derived one
possible density profile for such particles, assuming that equilibrium
is established between those undergoing capture, annhilation, and
escape from the Earth's gravitational potential.  We do not attempt to
fit our equation~(\ref{Sol_Dens2}) to this profile at all radii, but 
merely take the predicted neutrino density at the Earth's surface as
illustrative.  From Fig.~2 of Krauss et al. (1986), the expected 
escape rate for 10~GeV neutrinos is $2 \times 10^{16}$~s$^{-1}$, 
which translates into a density at the Earth's surface of
$\rho(R_{\oplus}) = 3 \times 10^{-20}$~kg m$^{-3}$ 
(about 50 times the canonical local halo dark matter density of 
$5 \times 10^{-22}$~kg m$^{-3}$).  If we suppose that this is 
rather associated with {\em solitonic\/} matter making up some 
fraction $\zeta$ of the Earth's total mass ($M_g=\zeta M_{\oplus}$),
then equation~(\ref{Sol_Dens2}) gives $b = -4 \times 10^{-14}\zeta^{-2}$.
For dark matter of this kind to be significant, $b$ must be small
for solar system bodies.  With $\zeta \sim 10^{-3}$, for example,
we have $b \sim -4 \times 10^{-8}$, while $\zeta \sim 10^{-6}$ would
correspond to $b \sim -0.04$.  These numbers are consistent with the
experimental limits obtained in \S\S~\ref{sec:LD}~-~\ref{sec:GE}
above.  It may be possible to constrain the theory more tightly by 
looking at violations of the {\em weak equivalence principle\/} by
solar system bodies (\cite{Ove00}).

On larger scales, systems such as galaxies and clusters of galaxies 
are suspected by many to harbor significant amounts of relativistic
hot dark matter.  
We take here as an example a recent numerical simulation (\cite{Kof96})
in which light (2.3~eV) neutrinos make up 20\% (by mass) of a cluster 
whose total mass $M_{\scriptscriptstyle T} = 6 \times 10^{14} M_{\odot}$.
Fig.~3 of this paper shows a typical neutrino density of 
$\rho(r) \approx 200 \rho_c$ at $r = 0.03$~Mpc, where
$\rho_c = 2 \times 10^{-26} h_0^2$~kg m$^{-3}$ is the critical density.
If this were instead attributed to {\em solitonic\/} dark matter of
total mass $M_g=\zeta M_{\scriptscriptstyle T}$, then the latter would
have $b = -0.01 \zeta^{-2}$ by equation~(\ref{Sol_Dens2}), where we have 
taken $h_0=0.65$.  If {\em all\/} the hot dark matter were solitonic
($\zeta = 0.2$), then $|b|$ could be as large as 0.3.
These values are illustrative only, since density profiles of hot 
dark matter in clusters are likely somewhat shallower than that 
indicated by equation~(\ref{Sol_Dens2})\footnote{
  Density profiles with $\rho \propto r^{-4}$ at large $r$ have
  however been discussed in other contexts, such as elliptical
  galaxies (\cite{Jaf83,deZ85,Her90}).}.
Nevertheless they establish that values of $|b|$ in galaxy clusters
might in principle be significantly larger than those in the solar
system, and this encourages us to speculate that stronger tests of
higher-dimensional gravity might be carried out using the excellent
observational data now available on gravitational lensing by 
these objects.

\section{Conclusions}

We have re-examined the classical tests of general relativity, as well
as the geodetic precession test, when Einstein's theory is extended from
four to five dimensions.  The physical meaning of previous calculations 
for light deflection and time delay have  been clarified physically, 
and the restriction of zero momentum and/or spin along the extra 
coordinate that characterized the earlier calculations of 
perihelion shift and geodetic precession has been lifted.

Our results show that Kaluza-Klein gravity remains consistent with
experiment.  The free parameter of the theory, however, is increasingly 
constrained to small values.  Thus, data on light deflection, radar 
ranging to Mars and the perihelion precession of Mercury all imply
a value of $|b| \leq 0.07$ for the Sun.  Improved data on solar
oblateness should improve the sensitivity of the perihelion precession
bound by as much as an order of magnitude.  And the upcoming launch of
Gravity Probe~B will allow us to measure values of $|b|$ for the Earth 
with an accuracy of one part in $10^4$ or better.

\acknowledgments

We thank P.~S.~Wesson for comments.  H.~L.~acknowledges the support of
the National Natural Science Foundation of China (grant no. 19975007).
J.~O. thanks R.~I.~Bush and W.-T.~Ni for comments on solar oblateness
and radar ranging, and acknowledges the support of the National Science 
and Engineering Research Council of Canada.  He also expresses his 
gratitude for the hospitality of C.~W.~F.~Everitt and the theory group
at Gravity Probe~B, Stanford University, where part of this work was
carried out.

\end{document}